**SUBJECT AREAS: PHYSICS, MATERIALS SCIENCE, CONDENSED-MATTER PHYSICS, SUPERCONDUCTING PROPERTIES AND MATERIALS**

Correspondence and requests for materials should be addressed to *K.K. (kudo@science.okayama-u.ac.jp) or **M.N. (nohara@science.okayama-u.ac.jp)

# Superconductivity in $Ca_{10}(Ir_4As_8)(Fe_2As_2)_5$ with Square-Planar Coordination of Iridium

Kazutaka Kudo,[1,*] Daisuke Mitsuoka,[1] Masaya Takasuga,[1] Yuki Sugiyama,[2] Kento Sugawara,[2] Naoyuki Katayama,[2] Hiroshi Sawa,[2] Hiroaki S. Kubo,[1] Kenta Takamori,[1] Masanori Ichioka,[1] Tatsuo Fujii,[3] Takashi Mizokawa,[4] and Minoru Nohara[1,**]

[1]Department of Physics, Okayama University, Okayama 700-8530, Japan

[2]Department of Applied Physics, Nagoya University, Nagoya 464-8603, Japan

[3]Department of Applied Chemistry and Biotechnology, Okayama University, Okayama 700-8530, Japan

[4]Department of Complexity Science and Engineering & Department of Physics, The University of Tokyo, Kashiwa 277-8561, Japan

**We report the unprecedented square-planar coordination of iridium in the iron iridium arsenide $Ca_{10}(Ir_4As_8)(Fe_2As_2)_5$. This material experiences superconductivity at 16 K. X-ray photoemission spectroscopy and first-principles band calculation suggest Ir(II) oxidation state, which yields electrically conductive $Ir_4As_8$ layers. Such metallic spacer layers are thought to enhance the interlayer coupling of $Fe_2As_2$, in which superconductivity emerges, thus offering a way to control the superconducting transition temperature.**

Platinum exhibits a rich variety of coordination geometries. For instance, all of the basic polyhedral forms, including octahedral,[1] triangle-planar,[2] tetrahedral,[3] and square-planar,[4–7] can be seen in platinum arsenides. The diversity of coordination chemistry allows us to synthesize many functional materials, such as superconductors. The following are prominent platinum-arsenide superconductors: $SrPt_2As_2$, which consists of $PtAs_4$ tetrahedra,[3] exhibits superconductivity at a transition temperature of $T_c$ = 5.2 K,[8] in which a charge transfer from donor to acceptor layers[9] and subsequent emergence of charge-density waves has been discussed;[3,8] SrPtAs, which consists of $PtAs_3$ triangles,[2] shows superconductivity at 2.4 K,[10] for which a broken time-reversal symmetry in a locally noncentrosymmetric structure has been proposed;[11] $Ca_{10}(Pt_4As_8)(Fe_{2-x}Pt_xAs_2)_5$, which consists of $PtAs_4$ planar squares, exhibits superconductivity at 38 K,[4–7] and therefore constitutes a member of the iron-based superconductors.[12–14] Palladium exhibits similar coordination chemistry;[15–18] $Ca_{10}(Pd_3As_8)(Fe_{2-x}Pd_xAs_2)_5$ with $PdAs_4$ planar squares was reported very recently to exhibit superconductivity at 17 K.[18]

In contrast, iridium shows limited coordination geometries; only octahedral and tetrahedral coordination are known in arsenides, as in $IrAs_3$[19] and $SrIr_2As_2$.[3] In this paper, we report the occurrence of square-planar coordination of iridium in a novel iron iridium arsenide $Ca_{10}(Ir_4As_8)(Fe_2As_2)_5$. This is the first inorganic compound that includes square-planar coordination of iridium. This compound exhibits superconductivity at $T_c$ = 16 K. First-principles calculations and X-ray photoelectron spectroscopy (XPS) suggest the presence of iridium (II) oxidation state. The resultant metallic nature of $Ir_4As_8$ spacer layers will be discussed.

## Results

### Crystal structure.

Single-crystal structure analysis revealed that the compound, discovered in this study, crystallizes in the tetragonal structure with the space group *P*4/*n* (#85) with a chemical composition of $Ca_{10}(Ir_4As_8)(Fe_2As_2)_5$ (see the Supplementary Tables S1 and S2 for crystallographic data) (CCDC 962099). The atomic ratios of Ca:Fe:Ir:As = 10:10:4:18 are consistent with the results of energy dispersive X-ray spectrometry, 10:9.8:5.8:20.1. The structure consists of alternating stacking of $(Fe_2As_2)_5$ and $Ir_4As_8$ layers with five Ca ions between them, as shown in Figure 1. This is isotypic to $Ca_{10}(Pt_4As_8)(Fe_{2-x}Pt_xAs_2)_5$[6] or $\alpha$-$(CaFe_{1-x}Pt_xAs)_{10}Pt_{4-y}As_8$.[7] The $Fe_2As_2$ layers, composed of edge-sharing $FeAs_4$ tetrahedra, are the common building block among iron-based superconductors.[12–14] The $Ir_4As_8$ layers are unique to the present compound, and act as spacer layers. The size of the Ir square lattice (with an Ir-Ir distance of 4.411 Å) is larger than that of the $Fe_2As_2$ square lattice (3.860-3.924 Å). This lattice mismatch leads to the formation of the $\sqrt{5} \times \sqrt{5}$ superstructure in the *ab*-plane, as shown in Figure 1c.

The characteristic square-planar coordination of Ir was found in the $Ir_4As_8$ layers. There are two Ir sites, as shown in Figure 1b. Ir1 adopts square-planar coordination, resulting in coplanar $IrAs_4$ squares with a Ir1-As3 bond length of 2.414 Å. On the other hand, Ir2 is at a non-coplanar site with respect to the $As_4$ square; Ir2 is displaced upward/downward by 0.676 Å toward the As4 ion at the adjacent $Fe_2As_2$ layer, as shown in Figure 1a. However, the distance between Ir2 and As4 (3.000 Å) is by far longer than the Ir2-As3 bond length (2.441 Å), thus Ir2 can be regarded as adopting square-planer coordination. The corner-sharing $Ir1As3_4$ and $Ir2As3_4$ squares constitute $Ir_4As_8$ layers, as shown in Figure 1b, where the As3 atoms form $As_2$ dimers with an As-As bond length of 2.454 Å, which comparable to twice the covalent radius of arsenic that is 2.42 Å.[18] These bond lengths are similar to those in platinum analogue, $Ca_{10}(Pt_4As_8)(Fe_{2-x}Pt_xAs_2)_5$[7]: Corresponding distances, Pt1-As3 = 2.484 Å, Pt2-As4 = 3.087 Å, and Pt2-As3 = 2.415 Å, suggest that the valence state of Ir is similar to that of Pt.

**Superconductivity.**

Figure 2 shows the temperature dependence of the in-plane electrical resistivity $\rho_{ab}$ of $Ca_{10}(Ir_4As_8)(Fe_2As_2)_5$. $\rho_{ab}(T)$ decreases with decreasing temperature, and shows a kink at approximately 100 K. This kink is not due to antiferromagnetic ordering, which is widely observed in iron-based superconductors,[12–14] since the single-peak structure of the $^{57}$Fe-Mössbauer spectrum at 300 K remains unchanged down to 50 K, as shown in the upper inset of Figure 2. At low temperatures, $\rho_{ab}(T)$ exhibits a sharp drop below 20 K, the characteristic of the onset of superconductivity. Zero resistivity was observed below 17 K. The 10–90% transition width was estimated to be approximately 2 K. The bulk superconductivity was evidenced by the temperature dependence of the magnetization *M*, shown in Figure 3. *M*(*T*) exhibits diamagnetic behavior below 16 K. The shielding signal estimated at 5 K corresponds to 83% of perfect diamagnetism.

## Discussion

The observed $T_c$ of 16 K is relatively low among iron-based superconductors.[12–14] We suggest that $Ca_{10}(Ir_4As_8)(Fe_2As_2)_5$ is in an overdoped region. The lower inset of Figure 2 shows the temperature dependence of the Hall coefficient $R_H$. The negative value suggests that the major carriers are electrons. The small value of $R_H$ as well as the small temperature dependence indicates the overdoping, as inferred from the $R_H$ of $Ba(Fe_{1-x}Co_x)_2As_2$.[20] This is consistent with the absence of antiferromagnetic ordering, which is characteristic of underdoped regions.[12–14] The consideration of charge neutrality based on the Zintl concept results in the same consequence. Assuming a divalent $Ir^{2+}$, the present compound is written as $Ca^{2+}{}_{10}(Ir^{2+}{}_4(As_2)^{4-}{}_4)(Fe^{2+}{}_2As^{3-}{}_2)_5 \cdot 2e^-$; the excess charge $0.2e^-$/Fe is intrinsically injected into the superconducting $Fe_2As_2$ layers. This doping level corresponds to overdoping, judging from the data on doped $BaFe_2As_2$.[21] We expect that a higher $T_c$ can be realized by reducing the intrinsic charge carriers.

Iron-based superconductors reported to date can be characterized by the insulating spacer layers,[12–14] which include rare-earth oxides[22] and alkaline-earth fluorides[23] with a fluorite-type structure, alkali[24] or alkali-earth[25] ion, and complex metal oxides with combined rock-salt and perovskite-type structures.[26–30] The insulating spacer layers are stacked in an alternating fashion with superconductive $Fe_2As_2$ layers, resulting in two-dimensional electronic Fermi surfaces that have been thought to be a key ingredient of high $T_c$ superconductivity.[12–14] In contrast, the $Ir_4As_8$ spacer layers of the present compound can be metallic: Figure 4 shows the partial density of states (pDOS) projections of Fe 3*d* and Ir 5*d* of $Ca_{10}(Ir_4As_8)(Fe_2As_2)_5$ from first-principles calculations using the WIEN2k package.[31] Fe 3*d* predominates in the pDOS at the Fermi energy ($E_F$), in common with the other iron-based superconductors.[32] A remarkable difference is noticeable in the pDOS of the spacer layers; a finite contribution of Ir 5*d* can be seen in the pDOS at $E_F$, suggesting that the $Ir_4As_8$ spacer layers are metallic. This is in contrast with the negligible pDOS at $E_F$ of the spacer layers for the other iron-based superconductors,[12–14,32] including the platinum analogue $Ca_{10}(Pt_4As_8)(Fe_2As_2)_5$: The $Pt_4As_8$ spacer layers are semiconducting because of the opening of the gap in the pDOS of Pt 5*d* at $E_F$.[7,33] The difference between the $Pt_4As_8$ and $Ir_4As_8$ layers might be attributed to that of the electron configurations; $Pt^{2+}$ ($5d^8$) forms a closed-shell configuration with a completely filled $d_{xy}$ orbital in the square-planar coordination, whereas $d_{xy}$ of $Ir^{2+}$ ($5d^7$) is formally half-filled, resulting in a metallic nature. The oxidation state of iridium (II) is suggested by first-principles calculations, which give an estimate of the total number of electrons of Ir1 and Ir2 (and thus the nominal oxidation states) to be 74.89 ($Ir^{2.11+}$) and 74.91 ($Ir^{2.09+}$) from the sum of pDOS up to $E_F$, respectively. This is consistent with XPS results, as shown in Figure 5: The binding energy at the peak position of Ir $4f_{7/2}$ spectrum suggests that the valence of Ir in $Ca_{10}(Ir_4As_8)(Fe_2As_2)_5$ is close to 2+, if we refer to the binding energy of $Ca_3CoIrO_6$[34] with $Ir^{4+}$ and assume that the binding energy is decreased by approximately 1 eV when the valence is decreased by 1 as inferred from the XPS data of $K_3IrBr_6$ and $K_2IrBr_6$.

In cuprates, it has been suggested that the interlayer coupling of superconducting $CuO_2$ planes enhances $T_c$.[35] The metallic nature of the spacer layers of the present compound $Ca_{10}(Ir_4As_8)(Fe_2As_2)_5$ may give rise to an opportunity to engineer the interlayer coupling of superconducting $Fe_2As_2$ and to thus further enhance the superconducting transition temperature. To do so, we have to develop chemical methods of optimizing the carrier concentration of $Ca_{10}(Ir_4As_8)(Fe_2As_2)_5$.

The unusual square-planar coordination of $Fe^{2+}$ has been reported for the oxide $SrFeO_2$.[36] It has been discussed that strong hybridization or covalent nature between Fe 3*d* and O 2*p* orbitals for $Fe^{2+}$ in the square-planar coordination is the key ingredient for the stability of $SrFeO_2$.[37] Similar mechanism might be applicable to the formation of the square-planar coordination of $Ir^{2+}$ of $Ca_{10}(Ir_4As_8)(Fe_2As_2)_5$ because of the strong hybridization between Ir 5*d* and As 4*p* orbitals.

In summary, we found the square-planar coordination of iridium in the $Ir_4As_8$ layers of the iron iridium arsenide $Ca_{10}(Ir_4As_8)(Fe_2As_2)_5$. This finding provided a novel iron-based superconductor with $T_c$ = 16 K. The optimization of the metallic spacer layer might offer a way to further increase the superconducting transition temperature of iron-based materials.

## Methods

### Preparation and characterization of samples.

Single crystals of $Ca_{10}(Ir_4As_8)(Fe_2As_2)_5$ were grown by heating a mixture of Ca, FeAs, $IrAs_2$, and Ir powders in a ratio of Ca:Fe:Ir:As = 10:10:4:18 or 10:26:14:40. The mixture was placed in an alumina crucible and sealed in an evacuated quartz tube. The manipulation was carried out in a glove box filled with argon gas. The ampules were heated at 700°C for 3 h and then at 1100–1150°C for 10–40 h, after which they were quenched in ice water. The quenching procedure was found to be crucial to obtaining the $Ca_{10}(Ir_4As_8)(Fe_2As_2)_5$ phase. This process yielded $Ca_{10}(Ir_4As_8)(Fe_2As_2)_5$ together with a small amount of powder mixture of $CaFe_2As_2$ and $IrAs_2$. Plate-like single crystals of $Ca_{10}(Ir_4As_8)(Fe_2As_2)_5$ with typical dimensions of 0.5 × 0.5 × 0.02 $mm^3$ were separated from the mixture. The crystals were characterized by synchrotron radiation X-ray diffraction,[38] energy dispersive X-ray spectrometry, and conventional transmission Mössbauer spectroscopy with a $^{57}Co$/Rh source.

### Electrical resistivity and magnetization measurements.

The electrical resistivity (parallel to the *ab*-plane) and Hall coefficient were measured using the Quantum Design PPMS. Magnetization was measured using the Quantum Design MPMS.

**X-ray photoelectron spectroscopy (XPS) measurements.**

The single crystals were cleaved under the ultrahigh vacuum for the XPS measurements that were carried out using JEOL JPS9200 analyzer and a Mg K$\alpha$ source (1253.6 eV). The total energy resolution was set to about 1.0 eV. The binding energy was calibrated using the Au 4f core level of the gold reference sample.

## Acknowledgments

Part of this work was performed at the Advanced Science Research Center, Okayama University. It was partially supported by Grants-in-Aid for Scientific Research (A) (23244074) and (C) (25400372) from the Japan Society for the Promotion of Science (JSPS) and the Funding Program for World-Leading Innovative R&D on Science and Technology (FIRST Program) from the JSPS. The synchrotron radiation experiments performed at BL02B1 and BL02B2 of SPring-8 were supported by the Japan Synchrotron Radiation Research Institute (JASRI; Proposal No. 2012A0083, 2012B0083, 2013A0083, and 2013A1197).

## Author contributions

K.K. and M.N. conceived and planed the research. D.M., M.T., and K.K. synthesized single crystals. Y.S., K.S., N.K., and H.S. performed single-crystal structural analysis using synchrotron radiation X-ray diffraction. D.M. and K.K. measured electrical resistivity and magnetization. T.F. carried out Mössbauer spectroscopy. H.S.K, K.T., and M.I. conducted first-principles calculations. T.M. carried out X-ray photoelectron spectroscopy. K.K. and M.N. discussed the results and wrote the manuscript.

## Additional information

Supplementary information: Crystallographic data of $Ca_{10}(Ir_4As_8)(Fe_2As_2)_5$ is available at http://www.nature.com/scientificreports.

Accession Codes: The crystal structure of $Ca(Ir_4As_8)(Fe_2As_2)_5$ has been deposited at the Cambridge Crystallographic Data Centre (http://www.ccdc.cam.ac.uk). Deposition number is CCDC 962099.

Competing financial interests: The authors declare no competing financial interests.

## Figure caption

**Figure 1. Crystal structure of $Ca_{10}(Ir_4As_8)(Fe_2As_2)_5$ with tetragonal structure [space group *P*4/*n* (#85)].** The thick solid lines indicate the unit cell. (a), (b), and (c) show the schematic overviews, $Ir_4As_8$ layer, and $(Fe_2As_2)_5$ layer, respectively. The blue and dark-blue hatches in (b) indicate $IrAs_4$ squares with coplanar Ir1 and non-coplanar Ir2, respectively. The dashed ellipsoids in (b) represent $As_2$ dimers.

**Figure 2. Temperature dependence of the electrical resistivity $\rho_{ab}$ for $Ca_{10}(Ir_4As_8)(Fe_2As_2)_5$.** The upper inset shows $^{57}$Fe-Mössbauer spectra together with fitted curves. The lower inset shows the temperature dependence of the Hall coefficient $R_H$.

**Figure 3. Temperature dependence of dc magnetization *M* for $Ca_{10}(Ir_4As_8)(Fe_2As_2)_5$ at a magnetic field *H* of 10 Oe in the zero-field and field cooling conditions.**

**Figure 4. Electronic density of states (DOS) for $Ca_{10}(Ir_4As_8)(Fe_2As_2)_5$.** The partial DOS projections (pDOS) of Fe 3*d* and Ir 5*d* are shown. The inset shows the pDOS of Ir 5*d* in the vicinity of the Fermi level $E_F$.

**Figure 5. Ir 4*f* photoemission spectrum of $Ca_{10}(Ir_4As_8)(Fe_2As_2)_5$ taken at 300 K compared to those of $Ca_3CoIrO_6$ and $IrAs_2$.** Broken lines represent the expected peak positions of Ir $4f_{7/2}$ of $Ir^{4+}$, $Ir^{3+}$, and $Ir^{2+}$ for oxides.

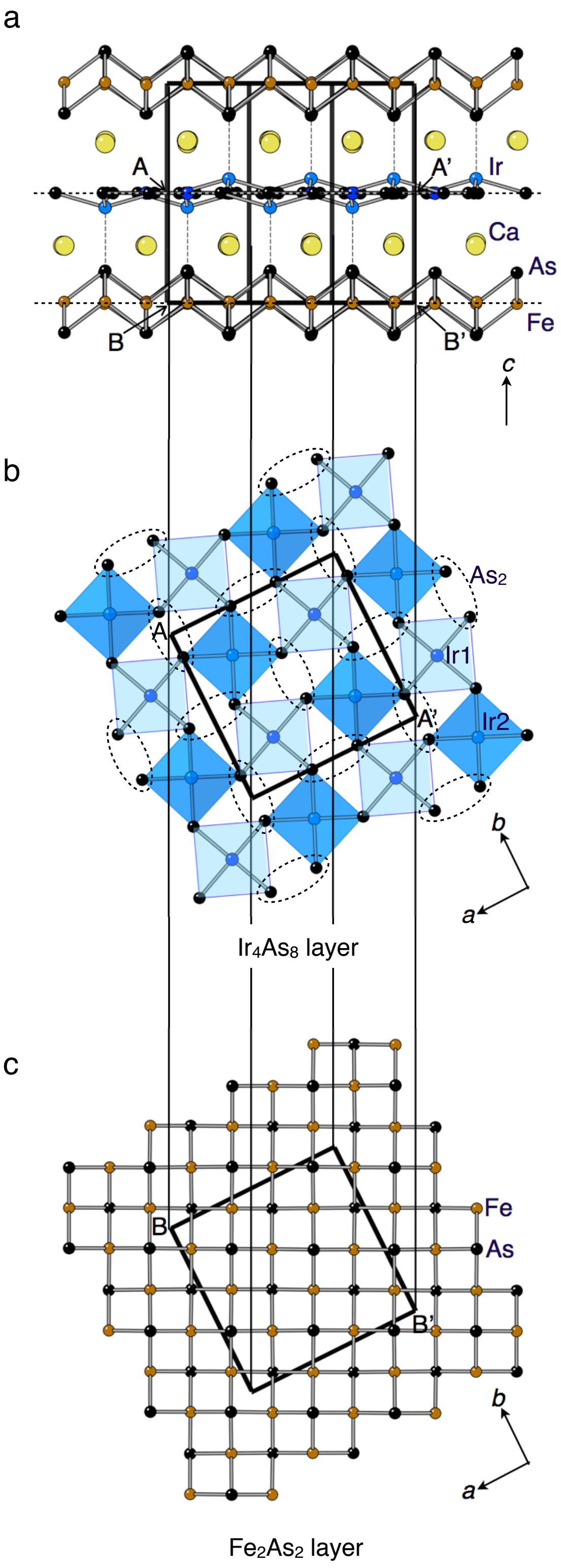

**Figure 1**

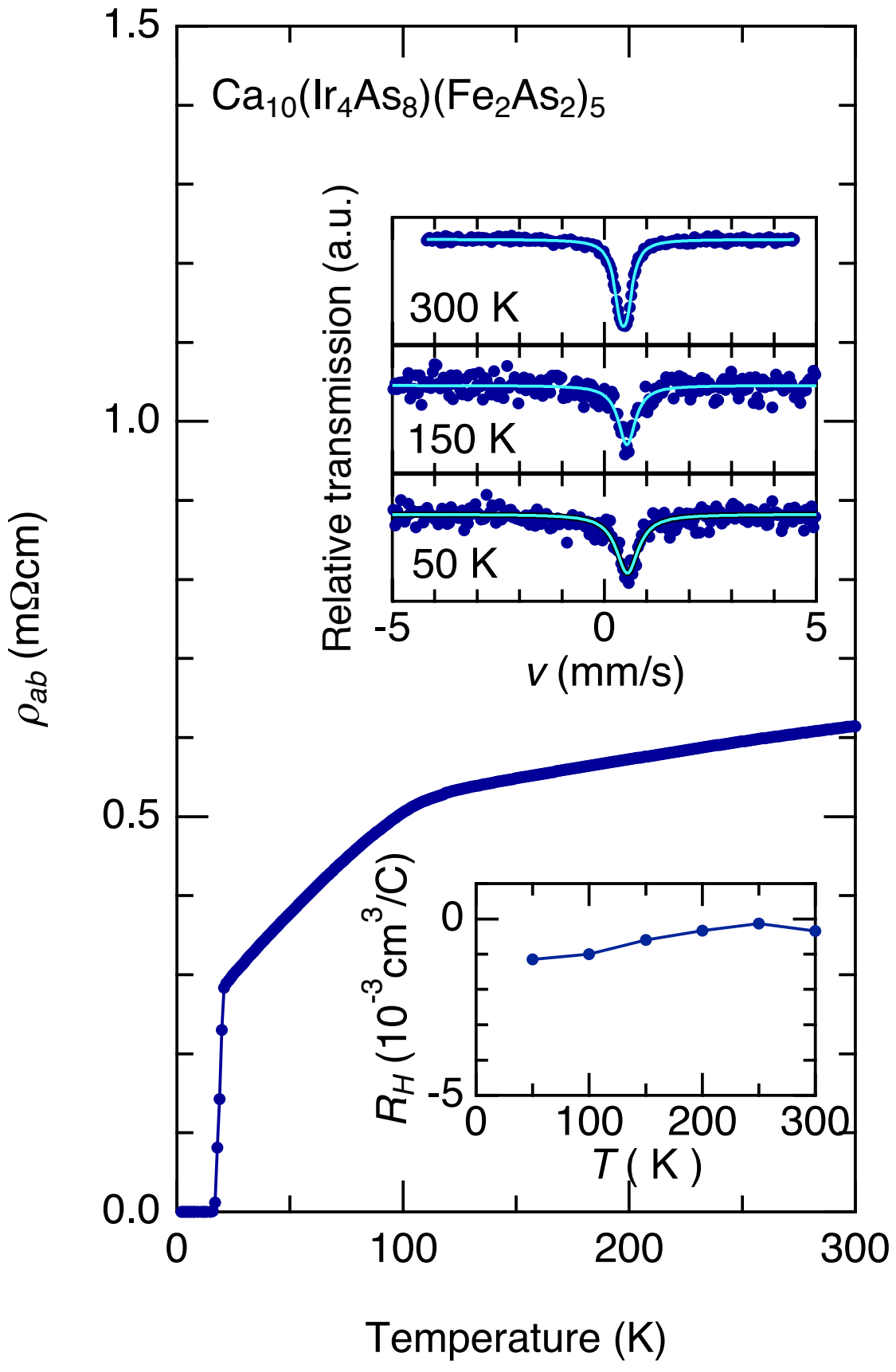

**Figure 2**

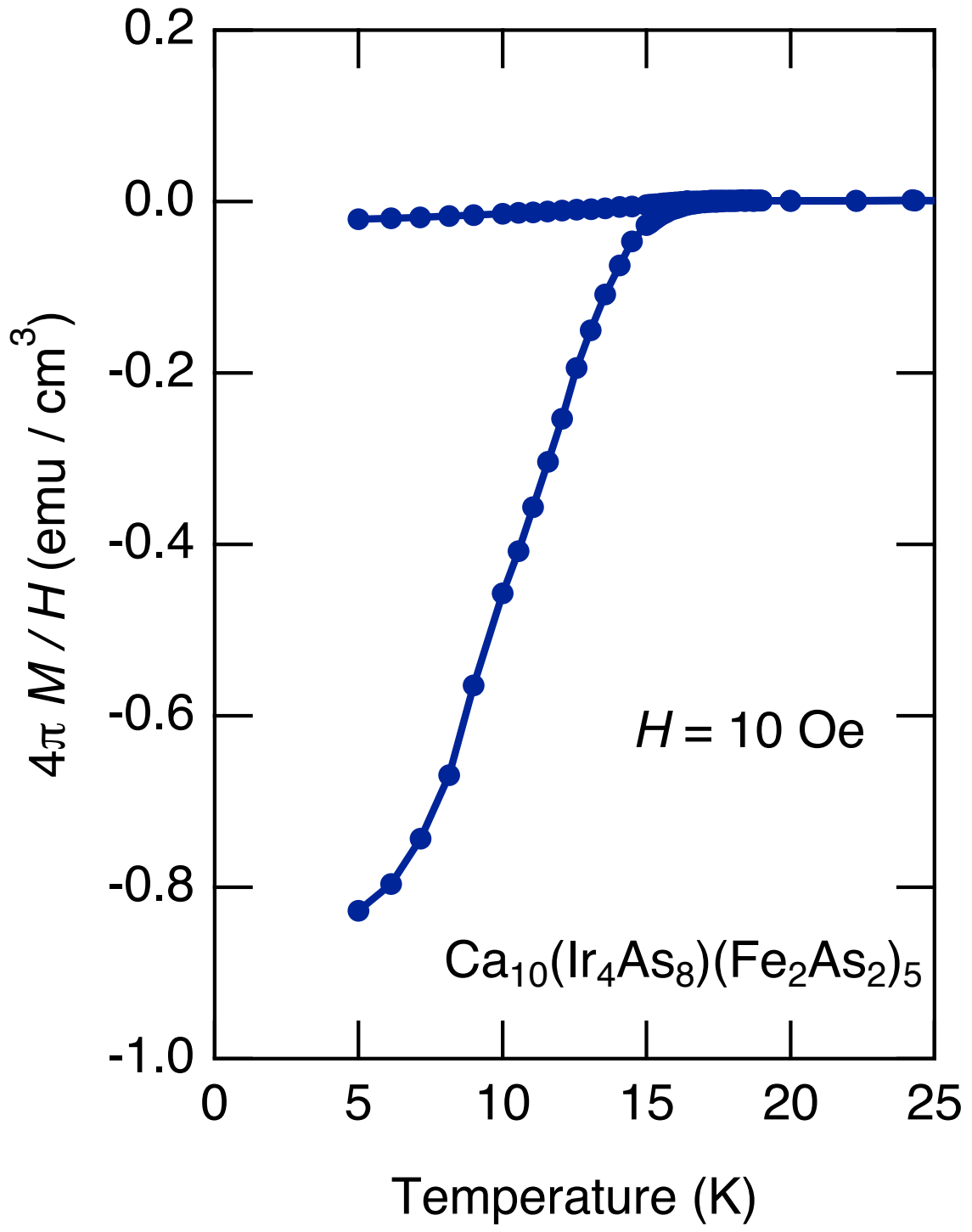

**Figure 3**

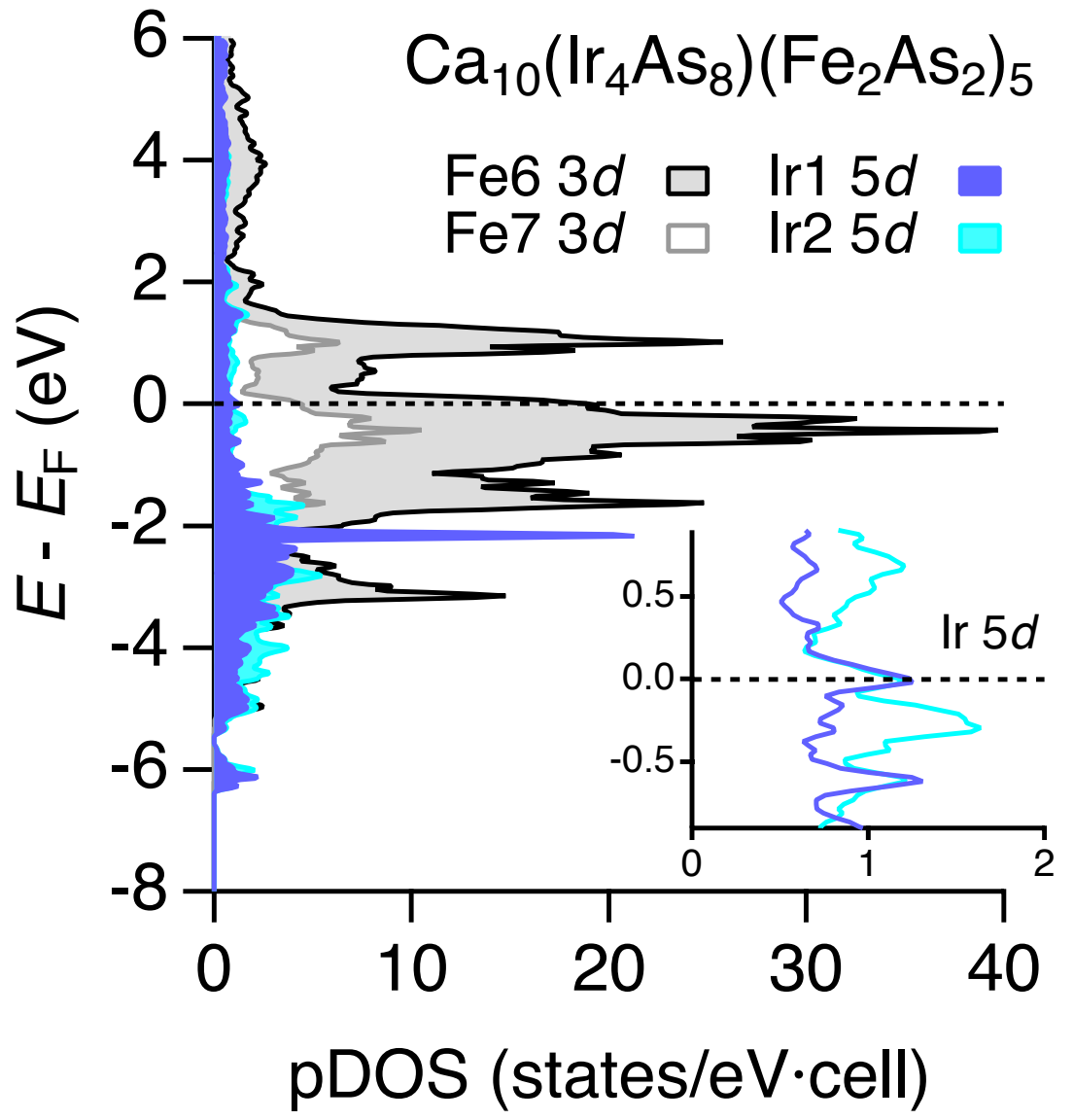

**Figure 4**

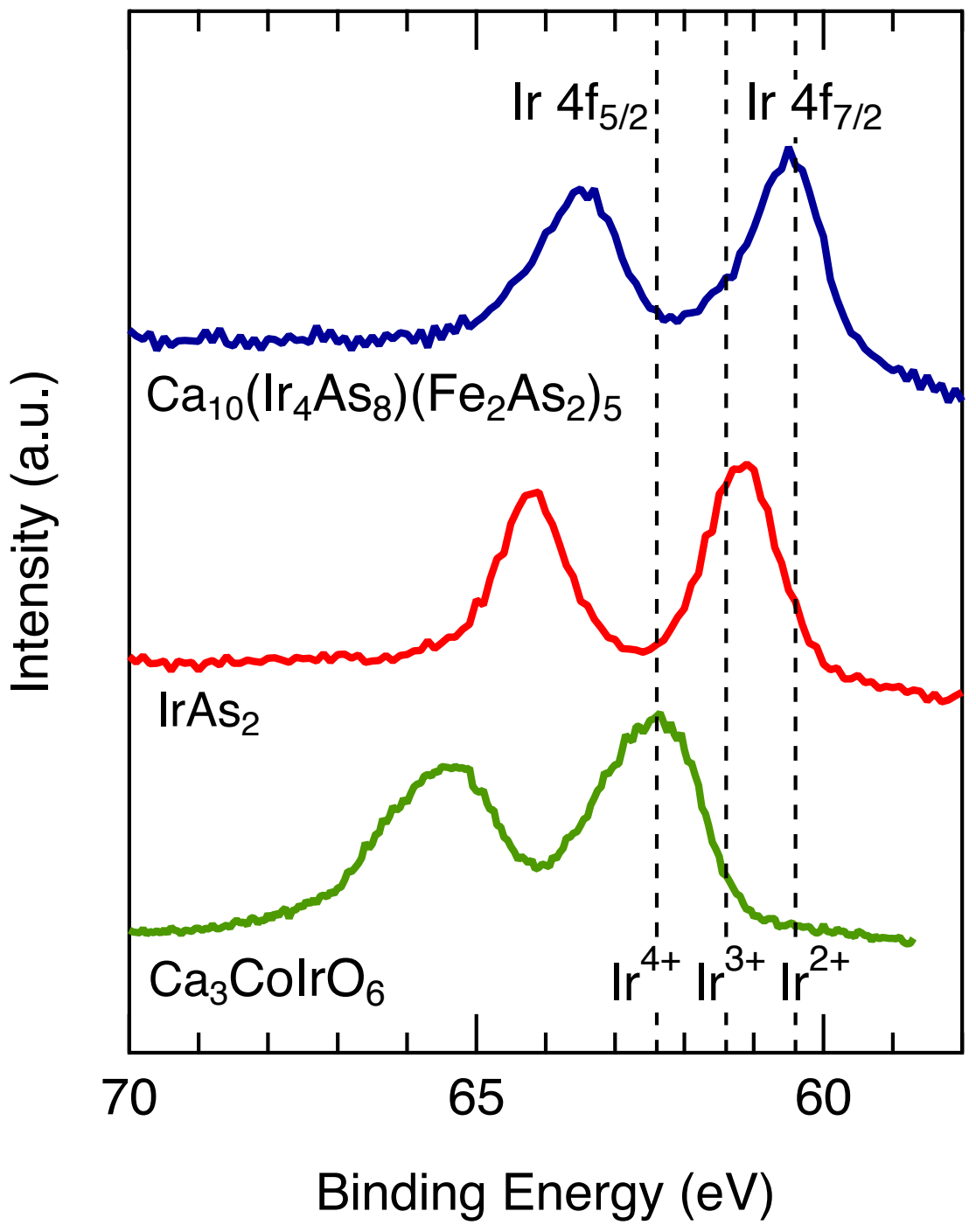

**Figure 5**

## Supplementary Information

# Superconductivity in $Ca_{10}(Ir_4As_8)(Fe_2As_2)_5$ with Square-Planar Coordination of Iridium

Kazutaka Kudo,[1] Daisuke Mitsuoka,[1] Masaya Takasuga,[1] Yuki Sugiyama,[2] Kento Sugawara,[2] Naoyuki Katayama,[2] Hiroshi Sawa,[2] Hiroaki S. Kubo,[1] Kenta Takamori,[1] Masanori Ichioka,[1] Tatsuo Fujii,[3] Takashi Mizokawa,[4] and Minoru Nohara[1]

[1]Department of Physics, Okayama University, Okayama 700-8530, Japan
[2]Department of Applied Physics, Nagoya University, Nagoya 464-8603, Japan
[3]Department of Applied Chemistry and Biotechnology, Okayama University, Okayama 700-8530, Japan
[4]Department of Complexity Science and Engineering & Department of Physics, The University of Tokyo, Kashiwa 277-8561, Japan

**Table S1.** Data collection and refinement statistics for the synchrotron X-ray structure determination of $Ca_{10}(Ir_4As_8)(Fe_2As_2)_5$.

| $Ca_{10}(Ir_4As_8)(Fe_2As_2)_5$ | 293(2) K |
|---|---|
| Data Collection | |
| Crystal System | tetragonal |
| Space Group | *P*4/*n* |
| *a* (Å) | 8.7198(3) |
| *b* (Å) | 8.7198(3) |
| *c* (Å) | 10.3768(12) |
| α, β, γ (°) | 90, 90, 90 |
| $R_{merge}$ | 0.0772 |
| *I* / σ I | > 2 |
| Completeness (%) | 0.982 |
| Redundancy | 14.2 |
| | |
| Refinement | |
| Resolution (Å) | 0.49 |
| No. of Reflections | 3476 |
| *R*1 | 0.0658 |
| No. of Atoms | 9 |

**Table S2.** Crystallographic parameters of $Ca_{10}(Ir_4As_8)(Fe_2As_2)_5$ with the space group *P*4/*n* at 293(2) K. The atomic coordinates and thermal parameters were refined. A crystal information file (CIF) of the crystal structure of $Ca_{10}(Ir_4As_8)(Fe_2As_2)_5$ derived by the analysis can be obtained free of charge from the Cambridge Crystallographic Data Centre at www.ccdc.cam.ac.uk. CCDC 962099 contains the crystallographic data for this paper.

| $Ca_{10}(Ir_4As_8)(Fe_2As_2)_5$ | | | | |
|---|---|---|---|---|
| | | Atomic Positions | | |
| Site | Occupancy | *x* / *a* | *y* / *b* | *z* / *c* |
| Ir1 | 1 | 1/4 | 3/4 | 1/2 |
| Ir2 | 1 | 3/4 | 3/4 | 0.56468(5) |
| As3 | 1 | 0.50426(7) | 0.64062(7) | 0.49949(5) |
| As4 | 1 | 1/4 | 1/4 | 0.14623(12) |
| As5 | 1 | 0.65104(6) | 0.45377(6) | 0.13589(6) |
| Fe6 | 1 | 0.44779(8) | 0.34930(9) | 0.00487(6) |
| Fe7 | 1 | 1/4 | 3/4 | 0 |
| Ca8 | 1 | 0.34543(13) | 0.55125(12) | 0.26048(12) |
| Ca9 | 1 | 3/4 | 3/4 | 0.2733(2) |